\documentclass[a4paper,11pt]{article}
\usepackage{pos}

\newcommand {\cH}{{\cal H}}

\newcommand {\cM}{{\cal M}}
\newcommand {\cN}{{\cal N}}
\newcommand {\cO}{{\cal O}}

\def\d{\delta}
\def\e{\epsilon}

\def\g{\gamma}

\def\l{\lambda}
\def\m{\mu}
\def\n{\nu}

\def\p{\pi}

\def\r{\rho}
\def\s{\sigma}

\def\L{\Lambda}

\def\U{\Upsilon}

\def\rd{{\rm d}}
\def\ri{{\rm i}}
\def\re{{\rm e}}

\newcommand{\ve}{\varepsilon}                            %

\newcommand{\pa}{\partial}                           %
\newcommand{\hf}{\frac12}

\newcommand{\vf}{\varphi}

\newcommand{\be}{\begin{equation}}
\newcommand{\ee}{\end{equation}}
\newcommand{\bea}{\begin{eqnarray}}
\newcommand{\eea}{\end{eqnarray}}

\newcommand{\bm}[1]{\mbox{\boldmath$#1$}}

\def\double #1{#1{\hbox{\kern-2pt $#1$}}}

\newif\ifdtup

\newcommand{\bsubeq}{\begin{subequations}}
\newcommand{\esubeq}{\end{subequations}}

\numberwithin{equation}{section}

\newcommand{\sU}{\mathsf{U}}

\title{Interacting $p$-form gauge theories: 
New developments}

\author
{Sergei M. Kuzenko}

\affiliation
{Department of Physics, The University of Western Australia,\\
  35 Stirling Highway, Perth W.A. 6009, Australia}

\emailAdd{sergei.kuzenko@uwa.edu.au}

\abstract{
Gauge $p$-forms in diverse dimensions are ubiquitous 
in supergravity and string theory. This work reviews novel covariant formulations designed 
to generate arbitrary interacting duality-invariant or chiral (self-dual) $p$-form theories in $d = 2p + 2$ space-time dimensions. For odd $p$, such theories possess $\mathsf{U}(1)$ duality invariance and include the Born-Infeld and ModMax theories. For even $p$, they describe a self-interacting chiral $p$-form with its gauge-invariant field strength obeying a nonlinear self-duality condition. We provide a complete description of $T\bar T$-like deformations of $\mathsf{U}(1)$ duality-invariant models for nonlinear electrodynamics in four dimensions and their six-dimensional counterparts -- interacting chiral two-form field theories. 
We also elaborate on consistent flows in the spaces of duality-invariant or chiral (self-dual) $p$-form theories
beyond six dimensions.}

\FullConference{The XVIth Quark Confinement and the Hadron Spectrum Conference (QCHSC24)\\
 19-24 August, 2024\\
 Cairns Convention Centre, Cairns, Queensland, Australia\\}

\begin{document}
\maketitle

\section{Introduction}

(Electromagnetic) duality and confinement are often interrelated, especially in supersymmetric Yang-Mills theories
\cite{Seiberg:1994rs}.
Patterns of duality invariance were observed in the late 1970s in extended supergravity
\cite{Ferrara:1976iq, Cremmer:1979up}, which 
have triggered renewed research into general aspects of duality invariance.\footnote{Renaissance of electromagnetic duality  (nonlinear self-duality) is one of many remarkable developments in quantum field theory and gravity inspired by the progress of supergravity. Others include the following: (i) a realisation of Einstein's dream of unifying gravity and electromagnetism that was achieved within the framework of $\cN=2$ supergravity \cite{FvN}; 
(ii) new types of gauge theories (as compared with Yang-Mills theories), with open gauge algebra and/or linearly dependent gauge generators (e.g., gauge $p$-forms in $d>p>1$ dimensions);
(iii) new covariant quantisation methods, including the Batalin-Vilkovisky formalism \cite{BV}; (iv) modern Kaluza-Klein theories, see \cite{Duff:1986hr} for a review; and (v) gauge/gravity duality, see \cite{Aharony:1999ti} for a review.
}  
This work reviews covariant models for a self-interacting gauge $p$-form in $2p+2$  dimensions such that (i) the equations of motion are $\sU(1)$ duality invariant for odd $p$; and (ii) the field strength satisfies a nonlinear self-duality condition on-shell for even $p$. We also discuss consistent flows in the space of 
$\sU(1)$ duality-invariant models for nonlinear electrodynamics and their six-dimensional counterparts -- interacting chiral two-form field theories. 
A remarkable feature of these deformations is that they are generated by functions of the energy-momentum tensor
($T\bar T$-like deformations), as was demonstrated in 
\cite{Ferko:2023wyi, Ferko:2024zth}.

Maxwell's electrodynamics in Minkowski space is the simplest and oldest example of a duality-invariant theory in four spacetime dimensions ($p=1$),
\bea
L_{\rm Maxwell}  (F)= - \frac 14 F^{\m\n} F_{\m\n} = \hf \big( \vec{E}^2 - \vec{B}^2 \big)~,
\qquad F_{\m\n } = \pa_\m  A_\n - \pa_\n  A_\m\, .
\eea
The Bianchi identity and the equation of motion are
\bea
\pa^\n \widetilde{F}_{\m\n} = 0\, , \qquad 
\pa^\n F_{\m\n} = 0 \, ,
\eea
with $\widetilde{F}^{\m\n} :=\hf  \ve^{\m\n \r \s}\, F_{\r\s} $  the Hodge dual of $F$. 
Since both differential equations have the same functional form, one may consider so-called duality rotations
\bea
F+\ri \widetilde F \to \re^{\ri \vf} \big( F+\ri \widetilde F \big) \quad \Longleftrightarrow \quad 
\vec{E} +\ri \vec{B} \to \re^{\ri \vf} \big( \vec{E} +\ri \vec{B} \big) ~, \qquad 
\vf \in {\mathbb R}\, .
\eea
 Lagrangian $L_{\rm Maxwell} (F)$ changes, but the energy-momentum tensor
\bea
T_{\m\n} = \hf \big( F+\ri \widetilde F\big)_{\m\r}  \big( F-\ri \widetilde F\big)_{\n\s} \eta^{\r\s}   
= F_{\m\r} F^{\n\s} \eta^{\r\s} - \frac 14 \eta_{\m\n} F^{\r\s} F_{\r\s} 
\eea
remains invariant under  these $\sU(1)$  duality transformations.
An extension of this simple linearised story to the nonlinear case was achieved in the mid-1990s. 
Building on the seminal work by Gaillard and Zumino \cite{Gaillard:1981rj}, 
the general formalism of $\sU(1)$ duality-invariant
models for nonlinear electrodynamics in four dimensions was developed in \cite{Gibbons:1995cv, Gibbons:1995ap, Gaillard:1997zr}.
 The Gaillard-Zumino-Gibbons-Rasheed  
(GZGR) approach \cite{Gaillard:1981rj, Gibbons:1995cv, Gibbons:1995ap, Gaillard:1997zr}
has been generalised to off-shell $\cN=1$ and $\cN=2$ globally \cite{KT1,KT2} and locally 
(see \cite{K13} and references therein) supersymmetric theories, 
as well as to higher-spin gauge models \cite{Kuzenko:2021qcx}.
In particular, the first consistent 
perturbative scheme to construct the $\cN=2$ supersymmetric Born-Infeld action 
was given in \cite{KT2}, and this approach was further pursued in \cite{BCFKR}. 

The GZGR formalism has also been extended to higher dimensions
\cite{Gibbons:1995cv, Tanii:1998px, Araki:1998nn, Aschieri:1999jr}.
In even dimensions, $d=2p+2$, 
the maximal duality group for a system of $k$ interacting gauge $p$-forms depends
on the dimension of spacetime. The duality group is U$(k)$ if $p$ is odd, 
and ${\rm O}(k) \times {\rm O}(k) $ if $p $ is even \cite{Araki:1998nn}.
So there is no interesting duality structure for a single ($k=1$) unconstrained gauge $2n$-form in $d =4n +2$ dimensions.
However, in $d =4n +2$ dimensions there exist gauge $2n$-forms with self-dual field strengths, also called chiral $2n$-forms, 
and the problem of constructing a covariant Lagrangian formulation for them is nontrivial even in the free case. 
It has been known since the mid-1980s \cite{Marcus:1982yu, Siegel:1983es}
that (i) a manifestly Lorentz invariant formulation for a chiral field is impossible without introduction of auxiliary fields; and 
(ii) if such a formulation exists, the Lagrangian should include at least cubic terms even in the free case.  
Several covariant formulations for chiral gauge fields have been proposed  in the literature, 
including those developed by Sen \cite{Sen:2015nph} and by Mkrtchyan and collaborators \cite{Mkrtchyan:2019opf}.
The most economic approach was pioneered by Pasti, Sorokin and Tonin (PST) \cite{Pasti:1995tn}. 
The PST formulation involves a single auxiliary scalar field, and it directly leads to the 
Hamiltonian-like formulation of \cite{Henneaux:1987hz, 
Schwarz:1993vs} upon imposing an appropriate gauge condition. 
In an important work \cite{Buratti:2019guq} Buratti, Lechner and Melotti extended the PST formalism to the case of general chiral form theories beyond six dimensions.

The PST approach has been used for various applications, including the construction of the M5-brane action of M-theory \cite{Bandos:1997ui} (see \cite{Sorokin:2025ezs} for a recent review). It
turns out to be well suited for the study of $T\bar T$-like flows in the space of self-interacting chiral two-form theories in six dimensions and their higher-dimensional analogues. 
Since 2016 there has been much interest in two-dimensional 
quantum field theories deformed by the irrelevant operator $T\bar T = \det (T_{\m\n})$  \cite{Smirnov:2016lqw, Cavaglia:2016oda}. Examples of deformations generated by the energy-momentum have also been discovered among effective gauge field theories beyond two dimensions, see e.g. \cite{Brizio:2024arr} for a review.  
Here we demonstrate, following \cite{Ferko:2023wyi, Ferko:2024zth},  the universality of $T\bar T$-like flows in the spaces of  $\sU(1)$ duality-invariant models for nonlinear electrodynamics in four dimensions and interacting chiral two-form field theories in six dimensions.  


\section{Interacting gauge $(2n-1)$-forms in $d=4n$ dimensions}

Different types of theories for gauge $p$-forms $A_{\m_1 \dots \m_{p}}$ in $d=2p +2$ dimensions arise depending on $p$ being even or odd, due to the property 
\bea
\widetilde{\widetilde{F}} = - (-1)^{d/2} F~, \qquad d= 2p+2
\label{Hodge}
\eea
of  the  Hodge dual, $\widetilde{F}_{\m_1 \dots \m_{p+1}}$,
 of the field strength $F_{\m_1 \dots \m_{p+1}}$.
In this section our attention is restricted to the case of odd $p$, and we relabel $d=2p +2 \to d=2p$.

\subsection{Gaillard-Zumino-Gibbons-Rasheed-type formalism}

Here we briefly sketch the extension of the GZGR formalism  to higher dimensions as described in 
\cite{Gibbons:1995cv, Tanii:1998px, Araki:1998nn, Aschieri:1999jr}.
In a curved spacetime  $\cM^d$  of even dimension $d=4n \equiv 2p$, with $n$ a positive integer,
we  consider
a self-interacting theory of a  gauge $(p-1)$-form $A_{\m_1 \dots \m_{p-1}}$
such that its Lagrangian, $L=L (F)$,
 is a function of the field strengths 
$F_{\m_1 \dots \m_p} =p  \pa_{ [\m_1} A_{\m_2 \dots \m_{p} ] }$.
In order for this theory to possess $\sU(1)$ duality invariance,
the Lagrangian must satisfy the self-duality equation \cite{Gibbons:1995cv,Araki:1998nn}
\bea
\widetilde G^{\m_1 \dots \m_p} G_{\m_1 \dots \m_p} 
+ \widetilde F^{\m_1 \dots \m_\p} F_{\m_1 \dots \m_p} =0~,
\label{self-duality}
\eea
where we have introduced\footnote{In this section, a partial derivative of $L(F)$ is defined by 
$ \d L(F) = 
\frac{1}{p!}
\d F_{\m_1 \dots \m_p} \frac{\pa L (F)}{\pa F_{\m_1 \dots \m_p}}$.}
\bea
\widetilde{G}^{\m_1 \dots \m_p} (F)= 
\frac{\pa L (F)}{\pa F_{\m_1 \dots \m_p}}~.
\eea
As usual, the notation $\widetilde F$ is used  for the Hodge dual of $F$, 
\bea
\widetilde{F}^{\m_1 \dots \m_p} =
\frac{1}{p!} \, {\bm \ve}^{\m_1 \dots \m_p \n_1 \dots \n_{p}} \,
F_{\n_1 \dots \n_{p} }~, \qquad {\bm \ve}^{\m_1 \dots \m_d} := \frac{1}{ \sqrt{-g}} \e^{\m_1 \dots \m_d}~,
\eea
with $\e^{\m_1 \dots \m_d}$ the ordinary Levi-Civita symbol.
Every solution of \eqref{self-duality} defines a  $\sU(1)$ duality-invariant theory.
An infinitesimal U(1) duality transformation is given by 
\bea
\d   \left( \begin{array}{c}  G \\ F \end{array} \right)
=  \left( \begin{array}{cr} 0 & 
- \l \\  \l  &  0 \end{array} \right) 
\left( \begin{array}{c}  G \\ F  \end{array} \right) ~,
\label{duality-tran}
\eea
with $\l$ a constant parameter. The simplest solution of \eqref{self-duality}  is the free Lagrangian
\bea
L_{\rm free} = -\frac{1}{2 p!}  F^{\m_1 \dots \m_p} F_{\m_1 \dots \m_p}~.
\eea


\subsection{Flows in the space of duality-invariant theories}

A function $\cO (F)$ is said to be a duality-invariant observable if it obeys the first-order differential equation
\bea
\frac{\pa \cO (F)}{\pa F_{\m_1 \dots \m_p}} G_{\m_1 \dots \m_p} =0~.
\eea
In accordance with \eqref{duality-tran}, $\cO(F)$ is inert under the duality transformations. An example of a duality-invariant observable is the energy-momentum tensor $T_{\m\n}$.
This follows from a simple generalization of the arguments given in  
\cite{Gaillard:1981rj, Gaillard:1997zr}.
Specifically, let $L(F; \g)$ be the Lagrangian of a $\sU(1)$ duality-invariant theory, where $\g$ is a duality-invariant parameter; $L(F; \g)$  is a solution of \eqref{self-duality} for every value of $\g$. 
Then, the function $\pa L(F; \g) / \pa \g$ is  duality invariant. 

Duality-invariant observables generate consistent flows in the space of field theories describing the dynamics of self-interacting gauge $p$-forms in the following sense
\cite{Ferko:2023wyi, Ferko:2024zth}. 
Let $L^{(\g)}  (F ) $ and $\cO^{ (\g) } ( F)$
be two scalar functions that depend on a real parameter $\g \in (-\e, \e)\subset \mathbb R$ and satisfy the following conditions:
\begin{enumerate} 
\item $L^{(\g)}$ and $\cO^{(\g)}$  obey the equations 
\bea
 \frac{\pa}{\pa \g} L^{(\g)} = \cO^{(\g)} \, ,\qquad 
 \frac{\pa \cO^{(\g)} (F)}{\pa F_{\m_1 \dots \m_p}} G^{(\g)}_{\m_1 \dots \m_p} =0~.
 \eea
\item  $L^{(0)}  ( F ) $ is a solution of \eqref{self-duality}. 
\end{enumerate}
Then $L^{(\g)}  ( F ) $ is a solution of \eqref{self-duality} at every value of the parameter $\g$.


\subsection{$T\bar T$-like flows in four dimensions}

Two important theorems concerning the $T\bar T$-like flows in four dimensions were proved in \cite{Ferko:2023wyi}. Consider a $\sU(1)$ duality-invariant nonlinear electrodynamics with Lagrangian $L(F)$.

{\bf Theorem 1.} 
Any two duality-invariant observables $\cO_1(F)$ and $\cO_2(F)$ prove to be functionally dependent, 
\bea
\U( \cO_1 , \cO_2) =0\, ,
\eea
for some function $\U$.

{\bf Theorem 2.}
 Every duality-invariant scalar observable $\cO(F) $ is a function of the energy-momentum tensor,  
 \bea
 \frac{\pa \cO (F)}{\pa F_{\m \n}} G_{\m \n} =0 \quad \implies \quad
 \cO = f(T_{\m\n})\, .
\eea

These theorems lead to the following important corollary. 
Given a one-parameter family of $\sU(1)$ duality-invariant theories, $L^{(\g)} (F)$,  Lagrangian obeys 
 $T\bar T$-like flow equation
\bea
\frac{\pa }{\pa \g}  L^{(\g)}   = \mathfrak{S}^{(\g)}  ( T_{\m\n} )\, .
\eea
Therefore, all consistent flows in the space of duality-invariant theories in $d=4$ are generated by the energy-momentum tensor.\footnote{An extension of the above theorems to $d=4n>4$ dimensions  is not yet known.} 

It is instructive to give two simple examples. For the Born-Infeld theory
\begin{align}
\sqrt{-g} \,L_{\rm BI} (F) &= \frac{1}{\g} \left\{\sqrt{-g}  - \sqrt{ -\det\Big(g_{\m\n} + \sqrt{\g} F_{\m \n} \Big)} \right\}  
\end{align}
it holds that \cite{Conti:2018jho}
\bea
 \frac{\pa L_{\rm BI}}{\pa \g} = \frac 18 \Big( T^{\m\n} T_{\m\n} - \hf (g_{\m\n} T^{\m\n})^2\Big)\, .
 \label{213}
\eea
For the ModMax theory \cite{BLST}
\begin{align}
    {L}_{\text{MM}} = - \frac{1}{4}  F\cdot F \cosh ( \gamma )+ \frac{1}{4} 
    \sqrt{ ( F\cdot F)^2 + \left( F \cdot \widetilde{F} \right)^2 } \sinh ( \gamma ) 
\end{align}
one obtains \cite{Babaei-Aghbolagh:2022uij, Ferko:2022iru}
\begin{align}
    \frac{\partial L_{\text{MM}} }{\partial \gamma} 
      = \frac{1}{2} \sqrt{ T^{\m\n} T_{\m\n} } \, .
    \label{215}
\end{align}
Supersymmetric extensions of the flows \eqref{213} and \eqref{215}
were derived in \cite{Ferko:2019oyv, Ferko:2023ruw} using the supercurrent results of \cite{Kuzenko:2005wh}.


\subsection{Generating formulation}

Ref. \cite{Kuzenko:2019nlm} proposed a reformulation of the $\sU(1)$ duality-invariant theories reviewed above, which for $p=2$ is equivalent to the formulation pioneered by  Ivanov and Zupnik  \cite{Ivanov:2002ab}.\footnote{The Ivanov-Zupnik formulation may be viewed as  a reformulation of the GZGR approach.} 
In this approach, along with the field strength $F_{\m_1\dots \m_p}$, the Lagrangian $L(F, V)$  depends on an auxiliary unconstrained rank-$p$ antisymmetric tensor $V_{\m_1 \dots \m_p}$ and has the form
\bea
L(F,V) = \frac{1}{p!} \Big\{ \hf F \cdot F +  V \cdot V - 2 V \cdot F\Big\} 
+ L_{\rm int} (V) ~,
\label{first-order}
\eea
where we have denoted
$ V\cdot F:= V^{\m_1 \dots \m_p} F_{\m_1 \dots \m_p}$.
 The last term in \eqref{first-order},  $L_{\rm int} (V) $, is at least quartic 
 in $V_{\m_1 \dots \m_p}$ provided $L(F)$ is analytic in $F$.
It is assumed that the equation of motion for $V$, 
\bea
\frac{\pa}{\pa V^{\m_1 \dots \m_p} } L(F,V) =0~,
\eea
allows one to integrate out the auxiliary field $V$ to result with $L(F)$. 

The self-duality equation \eqref{self-duality} proves
to be equivalent to the following condition on the self-interaction in \eqref{first-order}
\bea
\widetilde V^{\m_1 \dots \m_p} \frac{\pa}{\pa V^{\m_1 \dots \m_p} } L_{\rm int} (V) =0~.
\eea
Introducing (anti) self-dual components of $V$, 
\bea
V_\pm^{\m_1\dots \m_p} = \hf \Big( V^{\m_1\dots \m_p}  
\pm \ri \widetilde V^{\m_1\dots\m_np} \Big) ~, \qquad 
\widetilde V_\pm = \mp\ri V_\pm ~,\qquad V = V_+ +V_-~,
\eea
the above condition turns into 
\bea
\Big( 
 V_+^{\m_1 \dots \m_p} \frac{\pa}{\pa V_+^{\m_1 \dots \m_p} } 
 - V_-^{\m_1 \dots \m_p} \frac{\pa}{\pa V_-^{\m_1 \dots \m_p} } \Big)L_{\rm int} (V_+, V_-) 
=0~.
 \eea
This means that $ L_{\rm int} (V_+, V_-) $ is invariant under $\sU(1)$ phase transformations, 
\bea
L_{\rm int} (\re^{\ri \vf}  V_+, \re^{-\ri \vf} V_-)  = L_{\rm int} (V_+, V_-) ~, \qquad 
\vf \in {\mathbb R}~.
\label{manifestU(1)}
\eea
It should be pointed out that the duality transformation \eqref{duality-tran}  acts on $V$ as
\bea
\d V = \l \widetilde V~.
\eea

In four dimensions, the most general solution to the condition \eqref{manifestU(1)}
 is given by 
\bea
 L_{\rm int} (V_+, V_-) = f (V_+ \cdot V_+ V_- \cdot V_-) ~, 
 \eea
 with $f(x)$ a real function 
of one variable.
Similar solutions exist in higher dimensions. However more general 
self-interactions are possible beyond four dimensions.

Description of consistent deformations of $\sU(1)$ duality-invariant self-interacting gauge $(p-1)$-forms
is simple in the above approach based on 
the  formulation  \eqref{first-order}. One considers a Lagrangian of the form
\bea
L^{(\g)}(F,V) = \frac{1}{p!} \Big\{ \hf F \cdot F +  V \cdot V - 2 V \cdot F\Big\} 
+ L^{(\g)}_{\rm int} (V) ~,
\label{first-order-deformed}
\eea
where $L^{(\g)}_{\rm int} (V) $ is required to obey the condition \eqref{manifestU(1)}.


\section{Interacting chiral $2n$-forms in $d=4n+2$ dimensions}

In this section we turn to the case of self-interacting chiral $2n$-forms in $d= 4n + 2\equiv 2p +2$ dimensions, with $n \in \mathbb N$. Since $p$ in \eqref{Hodge} is now even, gauge $p$-forms with self-dual field strengths can be consistently defined in such dimensions. 
We start by reviewing  the work 
\cite{Buratti:2019guq}, in which Buratti, Lechner and Melotti extended the PST formalism to general chiral form theories
beyond six dimensions. 

\subsection{PST-type formulation}

Let $A_{\mu ( p ) } = A_{\m_1 \dots \m_p}$ be a gauge $p$-form potential\footnote{We often make use of the condensed notation  $ T_{\mu ( k ) } $ and $T^{\m(k)}$ for rank-$k$ antisymmetric tensors $T_{\mu_1 \ldots \mu_k} = T_{[\mu_1 \ldots \mu_k]} $  and $T^{\mu_1 \ldots \mu_k} = T^{[\mu_1 \ldots \mu_k]} $, respectively.} on a  time orientable spacetime $\cM^d$ with metric $g_{\m\n}$, 
and
\begin{align}
    F_{\mu ( p + 1 ) } = ( p + 1 ) \partial_{[\mu_1} A_{\mu_2 \ldots \mu_{p+1} ] }
\end{align}
be the corresponding gauge-invariant field strength. 
We introduce 
a normalised timelike vector field $v^\mu$,
$
    v^\mu v_\mu = - 1$.
Making use of $v^\m$ allows us to associate with $ F_{\mu ( p + 1 ) } $
the electric field 
\begin{align} \label{electric}
    E_{\mu ( p ) } = F_{\mu_1 \ldots \mu_p \nu } v^\nu \,  \qquad 
    E_{\mu_1 \ldots \mu_{n-1} \s } v^\s = 0 \, ,
\end{align}
and the magnetic field
\begin{align}\label{B_constraint}
    B_{\mu ( p ) } = \widetilde{F}_{\mu_1 \ldots \mu_p \nu} v^\nu\, , \qquad
    B_{\mu_1 \ldots \mu_{p-1} \s } v^\s = 0 \, .
\end{align}
Here we have introduced the dual field strength 
\begin{align}
    \widetilde{F}^{\mu ( p + 1 ) } := \frac{1}{(p+1)!} {\bm \varepsilon}^{\mu_1 \ldots \mu_{p+1} \nu_1 \ldots \nu_{p+1}} F_{\nu_1 \ldots \nu_{p+1}} \, ,
    \qquad
    \widetilde{\widetilde{F}} = F \, .
\end{align}
The field strength $ F_{\mu ( p + 1 ) } $ and its dual $   \widetilde{F}^{\mu ( p + 1 ) } $
are now expressed as 
\begin{subequations}
\begin{align}\label{F_expansion}
    F_{\mu ( p + 1 ) } &= - ( p + 1 ) E_{[ \mu_1 \ldots \mu_{p}} v_{\mu_{p+1} ]} - \frac{1}{p!} {\bm \varepsilon}_{\mu_1 \ldots \mu_{p+1} \nu_1 \ldots \nu_{p+1}} B^{\nu_1 \ldots \nu_p} v^{\nu_{p+1}} \, ,\\
\label{F_tilde_identity}
    \widetilde{F}_{\mu ( p+1 ) } &= - ( p + 1 ) B_{[ \mu_1 \ldots \mu_p} v_{\mu_{p+1} ] } - \frac{1}{p!} {\bm \varepsilon}_{\mu_1 \ldots \mu_{p+1} \nu_1 \ldots \nu_{p+1}} E^{\nu_1 \ldots \nu_p} v^{\nu_{p+1}} \, .
\end{align}
\end{subequations}
These identities lead to
\begin{align}
    F \cdot F &:= F^{\mu_1 \ldots \mu_{p+1}} F_{\mu_1 \ldots \mu_{p+1}} 
    = ( p + 1 ) \left( B \cdot B - E \cdot E \right) \, .
\end{align}

The authors of \cite{Buratti:2019guq} postulated the action
\begin{align}\label{BLM_action}
    S [ A, a ] = 
       \int \rd^d x \sqrt{-g}
     \left[ \frac{1}{2 p!} E \cdot B - \cH ( B_{\m(p)}, g_{\m\n} ) \right] \, , \qquad 
     v_\mu = \frac{\partial_\mu a}{\sqrt{ - \partial a \cdot \partial a } } \, ,
\end{align}
where $a(x)$ is an auxiliary  scalar field, and the function $\cH ( B_{\m(d)},   g_{\m\n} )$ must satisfy certain conditions specified below. 
Associated with $\cH$ is its derivative $ H_{\mu (p) } = \frac{\partial \cH ( B, g ) }{\partial B^{\mu (p)} }$
defined by 
\begin{align} \label{H-der}
    \delta_B \cH ( B , g ) = \frac{1}{p!} \delta B^{\mu_1 \ldots \mu_p} H_{\mu_1 \ldots \mu_p} \, , \qquad 
     H_{\mu_1 \ldots \mu_{p-1} \nu} v^\nu = 0 \, .
\end{align}
The equation of motion for the gauge $p$-form $A_{\m(p)} $ proves to be 
\bea
\partial_{[\m_1} \left( \mathbb{E}_{\mu_2 \ldots \mu_{p+1}} v_{\m_{p+2} ]} \right) =0\,,
\qquad   \mathbb{E}_{\mu ( p ) } = E_{\mu ( p ) } - H_{\mu ( p ) } \, .
\label{AEoM}
\eea

It  may be shown that the action \eqref{BLM_action} is invariant under PST gauge transformations of the form 
\begin{align} \label{PSTGT1}
         \delta A_{\mu ( p ) } = p v_{[\mu_1} \psi_{\mu_2 \ldots \mu_p]} \, , \qquad
    \delta a = 0 \, .
\end{align}
The second type of PST gauge transformations is:
\begin{align} \label{PSTGT2}
         \delta A_{\mu ( p ) } = - \frac{\varphi}{\sqrt{- \partial a \cdot \partial a}} \mathbb{E}_{\mu ( p ) } \, , \qquad
    \delta a = \varphi \, 
\end{align}
The $\varphi$-variation of the action proves to 
 vanish if
\begin{align}
   {\bm  \varepsilon}^{\sigma \rho \mu_1 \ldots \mu_p \nu_1 \ldots \nu_p} 
    \left( B_{\mu_1 \ldots \mu_p} B_{\nu_1 \ldots \nu_p} - H_{\mu_1 \ldots \mu_p} H_{\nu_1 \ldots \nu_p} \right) v_\rho = 0 \, .
\end{align}
Since $B_{\mu ( p )}$ and $H_{\mu ( p ) }$ are orthogonal to $v^\mu$, 
the above condition is equivalent to the equation
\begin{align} \label{MasterEquation}
    B_{[\mu_1 \ldots \mu_p} B_{\mu_{p+1} \ldots \mu_{2p} ] } 
    = H_{[ \mu_1 \ldots \mu_p } H_{\mu_{p+1} \ldots \mu_{2p} ] } \, ,
\end{align}
which is the condition on  $\cH ( B_{\m(d)},   g_{\m\n} )$.
Under this condition, the equation of motion for $a$ is identically satisfied 
if the equation of motion for $A_{\m(p)}$, eq. \eqref{AEoM}, holds.

Every solution of the equation  \eqref{MasterEquation} 
generates a consistent model for a chiral gauge $p$-form.
The case of a free chiral $p$-form corresponds to 
\bea
\cH^{\rm free}_{\rm PST}
\left(B, g \right) = \frac{1}{2 p!} B_{\mu (p)} B^{\mu (p)} \, .
\eea

Gauge freedom  \eqref{PSTGT1}
may be fixed in such a way that  
 \eqref{AEoM}  turns into
 \bea
   E_{\mu ( p ) } = H_{\mu ( p ) } (B, g) \, , \qquad H_{\mu (p) } (B,g)= \frac{\partial \cH ( B, g ) }{\partial B^{\mu (p)} }\, .
\eea
This is the generalised self-duality condition, which reduces to $E_{\mu ( p ) } = B_{\mu ( p ) } $ in the free case.


\subsection{Flows in the space of chiral  $2n$-form theories  in $d=4n+2$ dimensions}

Consistent flows in the space of a self-interacting chiral $2n$-form in $d= 4n + 2$ dimensions are generated by physical observables \cite{Ferko:2024zth}.
A  scalar function $\cO(B_{\m(p)} ) $ is called a physical observable 
(the dependence  of $\cO$ on the metric is assumed)
if it obeys the first-order differential equation
\begin{align}
   \cO_{[ \mu_1 \ldots \mu_p } H_{\mu_{p+1} \ldots \mu_{2p} ] } =0 \, ,
   \label{PhysicalObservable}
\end{align}
where the partial derivative 
$ \cO_{\mu (p) } = \frac{\partial \cO ( B ) }{\partial B^{\mu (p)} }$ is
defined by 
\begin{align}
    \delta_B \cO ( B, g ) = \frac{1}{p!} \delta B^{\mu_1 \ldots \mu_p} \cO_{\mu_1 \ldots \mu_p} \,  , \qquad 
    \cO_{\mu_1 \ldots \mu_{p-1} \nu} v^\nu = 0 \, .
\end{align}
Equation \eqref{PhysicalObservable} implies that $\cO$ is independent of $v_\m$ on the mass shell. Equivalently   \eqref{PhysicalObservable} proves to be the condition of Lorentz invariance of $\cO$. 

The following result was derived in \cite{Ferko:2024zth}.
 Let $\cH^{(\g)}  ( B_{\m(p)} ) $ and $\cO^{ (\g) } ( B_{\m(p)})$
be two scalar functions that depend on a real parameter $\g$ belonging to an open interval $ (-\e, \e)\subset \mathbb R$ and satisfy the following conditions:
\begin{itemize} 
\item $\cH^{(\g)}$ and $\cO^{(\g)}$  obey the equations 
\bea
 \frac{\pa}{\pa \g} \cH^{(\g)} = \cO^{(\g)} \, ,\qquad 
   \cO^{(\g)}_{[ \mu_1 \ldots \mu_p } H^{(\g)}_{\mu_{p+1} \ldots \mu_{2p} ] } =0 \, ;
 \eea
\item  $\cH^{(0)}  ( B_{\m(p)} ) $ is a solution of \eqref{MasterEquation}. 
\end{itemize}
Then $\cH^{(\g)}  ( B_{\m(p)} ) $ is a solution of \eqref{MasterEquation} at every value of the parameter $\g$.


\subsection{$T\bar T$-like flows in the space of chiral two-form theories in six dimensions}

In the case of six-dimensional nonlinear chiral two-form gauge theories, consistent deformations are driven by  $T\bar T$-like flows generated by the energy-momentum tensor. The following results were established 
in \cite{Ferko:2024zth}.\footnote{The theorems concerning the $T\bar T$-like flows in four and six dimensions \cite{Ferko:2023wyi, Ferko:2024zth} were generalised to  two dimensions in \cite{EFMTM}.}
\begin{itemize}

\item Given two invariant observables $\cO_1$ and $\cO_2$, they turn out to be functionally dependent.

\item Every invariant observable $\cO$ proves to be a function of 
the energy-momentum tensor,  $\cO = f(T_{\m\n})$. 

\end{itemize}
An extension of these results to  $d=4n+2>6$ dimensions is not yet known.


\subsection{Generating formulation}

The formulation\footnote{This formulation is a generalisation of the four-dimensional 
approach of \cite{Ivanov:2014nya}.} proposed in \cite{Ferko:2024zth} involves 
a general self-dual field 
$\Lambda_{\mu ( p+ 1 ) }$, $\widetilde{\L} = \L$,
which plays the role of a Lagrange multiplier. The action is
\begin{align}
\label{Interaction}
    S [ A , a , \Lambda ] &= \frac{1}{p!} \int \left[ \frac{1}{2} E \cdot B - \frac{1}{2} B \cdot B + ( B + \lambda )\cdot  ( B + \lambda ) ] \right] 
    - \int   \mathcal{V} \left(  \L_{\m(p+1)} , g_{\m\n}\right) \, ,
\end{align}
where we have defined
\begin{align}
    \lambda_{\mu ( p ) } = \Lambda_{\mu_1 \ldots \mu_p \nu} v^\nu \, .
\end{align}
The action is invariant under the following generalisation  of
  \eqref{PSTGT1} 
  \begin{align} \label{PSTGT4}
       \delta A_{\mu ( p ) } = p v_{[\mu_1} \psi_{\mu_2 \ldots \mu_p]} \, , \qquad
    \delta a = 0 \, , \qquad \d \L_{\m(p)}=0\, .
\end{align}
It is also invariant under
a simple generalisation of the gauge transformation \eqref{PSTGT2} given by 
\begin{align} \label{PSTGT3}
    \delta A_{\mu ( p ) } = \frac{\varphi}{\sqrt{ - \partial a \cdot \partial a}} \left[ \mathfrak{E} + 2 \left( B + \lambda \right) \right]_{\mu(p)} \, , \qquad \delta a = \varphi \, , \qquad
    \d \L_{\m(p+1)} =0\, ,
\end{align}
where we have denoted 
\begin{align}
    \mathfrak{E}_{\mu ( p ) } = E_{\mu ( p ) } - B_{\mu ( p ) } \, .
\end{align}
No condition on the interaction $  \mathcal{V} \left(  \L_{\m(p+1)} , g_{\m\n}\right) $ is imposed by the gauge freedom.

The auxiliary self-dual field $\Lambda_{\mu ( p+ 1 ) }$ may be eliminated from \eqref{Interaction}
 with the aid of its algebraic equation of motion. This may be seen to lead to the original theory \eqref{BLM_action}. The main virtue of the formulation 
  \eqref{Interaction} proposed in \cite{Ferko:2024zth}, 
   is that the interaction $  \mathcal{V} \left(  \L_{\m(p+1)} , g_{\m\n}\right) $  is an arbitrary scalar function of $\Lambda_{\mu ( p+ 1 ) }$. Therefore, \eqref{Interaction} provides a generating formulation for the self-interacting chiral $p$-form in $d= 2p + 2$ dimensions, with even $p$. A similar arbitrary function of a self-dual 
 $(p+1)$-form was also used in the Lagrangian formulation  by Avetisyan, Evnin and Mkrtchyan
   \cite{Mkrtchyan:2019opf} based on a different set of auxiliary fields.


\acknowledgments
\noindent
It is my pleasure to acknowledge  Professor Anthony Williams of the University of Adelaide for the invitation to give a talk at the XVth Quark Confinement and the Hadron Spectrum conference. I thank Emmanouil Raptakis, Dmitri Sorokin and Gabriele Tartaglino-Mazzucchelli for comments on the manuscript. I am very grateful to my colleagues Christian Ferko, Kurt Lechner, Liam Smith, Dmitri Sorokin and Gabriele Tartaglino-Mazzucchelli for the fruitful and enjoyable collaboration on the projects \cite{Ferko:2023wyi, Ferko:2024zth}. 
This work was supported in part by the Australian Research Council, project No. DP230101629.

\end{document}